\documentclass{article}
\usepackage{epsfig,a4}

\def\cal{\mathcal}

\def\to{\rightarrow}

\begin{document}

\includegraphics[width=3.5cm]{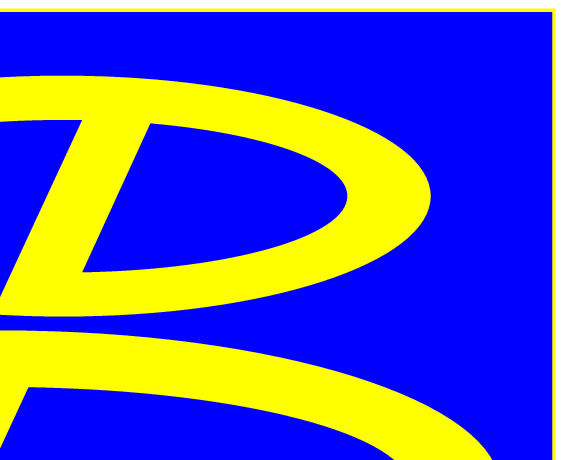}\\
\vspace*{-3.3cm}
\ \\
\hspace*{7.5cm} {\bf KEK Preprint 2003-61}\\
\hspace*{7.5cm} {\bf Belle Preprint 2003-19}\\
\hspace*{8.7cm} {\bf September 2003}\\
\hspace*{7.4cm} {\bf (Ver.2: November 2003)}\\
\ \\
\vspace{1.7cm}
 
%

\begin{center}
\Large
{\bf Measurement of  $K^+K^-$ production in two-photon\\ 
collisions in the resonant-mass region}

\normalsize

\vspace{0.5cm}




The Belle Collaboration\\
\ \\
   K.~Abe$^{5}$,               
   K.~Abe$^{35}$,              
   T.~Abe$^{5}$,               
   I.~Adachi$^{5}$,            
   H.~Aihara$^{37}$,           
   M.~Akatsu$^{17}$,           
   Y.~Asano$^{42}$,            
   T.~Aso$^{41}$,              
   V.~Aulchenko$^{1}$,         
   T.~Aushev$^{9}$
% ...
\end{center}

\small
\begin{center}
$^{1}${Budker Institute of Nuclear Physics, Novosibirsk}\\
% ...
$^{5}${High Energy Accelerator Research Organization (KEK), Tsukuba}\\
% ...
$^{9}${Institute for Theoretical and Experimental Physics, Moscow}\\
% ...
$^{17}${Nagoya University, Nagoya}\\
% ...
$^{35}${Tohoku Gakuin University, Tagajo}\\
% ...
$^{37}${Department of Physics, University of Tokyo, Tokyo}\\
% ...
$^{41}${Toyama National College of Maritime Technology, Toyama}\\
$^{42}${University of Tsukuba, Tsukuba}\\
% ...


\end{center}

\normalsize
\begin{abstract}
$K^+K^-$ production in two-photon collisions 
has been studied using a large data sample of 67~fb$^{-1}$ 
accumulated with the Belle detector at the KEKB asymmetric
$e^+e^-$ collider.
We have measured the cross section for the process
$\gamma\gamma\to K^+ K^-$ for center-of-mass energies
between 1.4 and 2.4 GeV, and found three new resonant structures  
in the energy region between 1.6 and 2.4~GeV.  
The angular differential cross sections have
also been measured. \\
\end{abstract}




\hspace*{18mm} -- 14.40.Cs Other mesons with $S=C=0$

\clearpage

\normalsize
\section{Introduction}
A high luminosity electron-positron collider is well suited 
for studies of
meson resonances produced in two-photon collisions. 
The heaviest established resonance that has 
been observed so far in kaon-pair production in 
two-photon processes is the $f'_2(1525)$ meson, which is classified as
an almost pure
$s\bar{s}$ meson. Above the $f'_2(1525)$ mass, no clear resonance structure
has been found in the $K^+K^-$ channel~\cite{cKK,cKK2}. 

 The L3 experiment at LEP has reported a 
resonance-like peak at 1.767~GeV in the
$\gamma\gamma \to K^0_SK^0_S$ process~\cite{L3}.  Their recent analysis shows
a dominant contribution of a tensor component in this energy region.
However, assignment of this structure to any known resonance state or
a spin/isospin state is not yet conclusive.  The corresponding resonant
structure is expected to appear also in the $K^+K^-$ channel because 
of the isospin invariance of resonance decay.  
However, we cannot predict its shape in the cross section
since two-photon reactions themselves are not isospin invariant
and the continuum contributions and interference effects
can differ between the $K^0 \bar{K^0}$ and $K^+K^-$ channels.
Therefore, measurements of the $\gamma\gamma \to K^+K^-$ and
$\gamma\gamma \to K^0_SK^0_S$ processes in the same mass region
give essentially independent information.

The 1.6 -- 2.4 GeV region is very important for meson spectroscopy since
radially-excited $q\bar{q}$ states are expected to exist in this mass range.
Several mesons with poorly measured properties are reported in this region
having $J^{PC} = {\rm (even)}^{++}$~\cite{pdg},
where $J$, $P$ and $C$ are spin,
parity and charge conjugation, respectively. The properties of
these mesons can be measured precisely using a large clean sample of 
$\gamma \gamma \to K \bar{K}$ events.

In addition, two-photon processes play an important role in identifying glueballs, since
we expect much smaller coupling of photons with a glueball
than with a $q\bar{q}$ meson.
Some glueball candidates around 2~GeV have been 
observed in $J/\psi \rightarrow \gamma
K\bar{K}$ decays~\cite{Jpsi}, 
but none has been identified conclusively as a glueball. 

Here, we report results on measurements of
$\gamma\gamma \rightarrow K^+K^-$ for two-photon center-of-mass 
energy between  1.4 and 2.4~GeV.
We describe the experimental apparatus and triggers in section~2.
In section~3, the selection criteria of the signal events are shown.
Section~4 introduces the methods used to derive the total and
differential cross sections and presents the results.
The major sources of systematic errors in the
present measurement are itemized in section~5.
We apply phenomenological analyses to the obtained
cross sections, mainly with respect to 
extractions of resonances in this energy region, in section~6,
and discuss the attributes of these resonant structures
in section~7. Conclusions are drawn in section~8.

\section{Experimental Data and the Belle detector}
The experiment is carried out with
the Belle detector~\cite{belle} at the KEKB asymmetric $e^+e^-$ 
collider~\cite{kekb}. In KEKB, the electron beam (8~GeV) and 
positron beam (3.5~GeV) collide with a crossing angle of 22~mrad.
The data collected between 1999 and April 2002 are used for this analysis,  
corresponding to an integrated 
luminosity of 67~fb$^{-1}$.
We combine data taken at the on- and off-resonance energies; 
the off-resonance data are taken 60~MeV below the $\Upsilon(4S)$ resonance 
at 10.58~GeV.

In the present analysis, neither the recoil electron nor positron is detected.
The basic topology of the signal events is 
just two charged tracks with opposite charge.  We
restrict the virtuality of the incident photons 
to be small by imposing a strict transverse-momentum balance requirement 
on this two-track system with respect to the
incident axis in the $e^+e^-$ center of mass (c.m.) frame.

  A comprehensive description of the Belle detector is
given elsewhere~\cite{belle}. We mention here only the
detector components essential for the present measurement.

Charged tracks are reconstructed from hit information in a central
drift chamber (CDC) located in a uniform 1.5~T solenoidal magnetic field.
The $z$ axis of the detector and the solenoid are along the positron beam,
with the positrons moving in the $-z$ direction.  The CDC measures the
longitudinal and transverse momentum components (along the $z$ axis and
in the $r\varphi$ plane, respectively).
The transverse momentum resolution is determined from cosmic 
rays and $e^+e^- \to \mu^+\mu^-$ events to be
$(\sigma_{p_t}/{p_t})^2 = 0.0030^2 + (0.0019p_t)^2$, where $p_t$ is the transverse 
momentum in GeV/$c$.
Track trajectory coordinates near the
collision point are provided by a
silicon vertex detector (SVD).  Photon detection and
energy measurements are performed with a CsI electromagnetic
calorimeter (ECL). In this analysis, the ECL is mainly used for 
rejection of electrons.
Identification of kaons is made using information from the
time-of-flight counters (TOF) and silica-aerogel Cherenkov
counters (ACC).  The ACC 
provides good separation between kaons and pions or muons at momenta above
1.2~GeV/$c$.    The TOF system consists of a barrel of 128 plastic scintillation
counters, and is effective in $K/\pi$ 
separation mainly for tracks with momentum below 1.2~GeV/$c$. 
The lower energy kaons are identified also using specific
ionization ($dE/dx$) measurements in the CDC.
The magnet return yoke is 
instrumented to form the $K_L$ and muon detector (KLM), 
which detects muon tracks and provides trigger signals.
In this analysis, 
the KLM is used only for the calibration of the trigger 
efficiencies using muon-pair events; it is not needed nor used 
for particle identification of the signal events.

Signal events are triggered most effectively by requiring two or more
CDC tracks in the $r\varphi$ plane
as well as two or
more TOF hits and at least one isolated cluster in the ECL with an energy above 0.1~GeV.
The opening angle in the $r\varphi$ plane of the two tracks must exceed 135$^\circ$,
and at least one of them must have $z$ coordinate information from the CDC cathodes.
Additional triggers that combine the two-or-more track requirement with
either a KLM track or an ECL energy deposit above 0.5~GeV also collect
some signal events; they are used in conjunction with other redundant triggers to
detect muon-pair and electron-pair events from two-photon collisions and thereby calibrate
the trigger efficiency for our signal events.

The trigger efficiency for the $K^+K^-$ process is 80-93\% in most 
of our selection acceptance, with the variation depending on
the transverse momentum of the tracks.


\section{Event Selection}

Candidate events are selected offline using the following criteria.
The event must have two oppositely charged tracks,
each with $p_t > 0.4$~GeV/$c$, $|dr| < 1$~cm, $|dz| < 3$~cm, and
$-0.47 < \cos \theta <+0.82$, and both satisfying  $|dz_1 - dz_2| \le 1$~cm.
Here, $p_t$ is the transverse momentum with respect to the positron-beam axis, 
$|dr|$ and $dz$ are the radial and axial coordinates, respectively,
of the point of closest approach to the nominal collision point 
(as seen in the $r\varphi$ plane), and $\theta$ is the polar angle, all measured
in the laboratory frame.  We demand that the event have no extra
charged track with $p_t > 0.1$~GeV/$c$, and then use only the two tracks with
$p_t > 0.4$~GeV/$c$ in this analysis.
The sum of the track momentum magnitudes must be less than 6~GeV/$c$, and the
total energy deposition in the ECL must be less than 6~GeV.

Events with an initial-state radiation, such as radiative Bhabha events,
are suppressed by requiring that the invariant mass of
the two tracks be smaller 
than 4.5~GeV/$c^2$ and the square of the missing mass of the event be greater than
2~GeV$^2/c^4$.
Cosmic ray events are rejected by requiring the cosine of the 
opening angle of the tracks to be 
greater than $-0.997$. 
Exclusive two-track events from quasi-real two-photon collisions are selected by
requiring a good transverse momentum balance in the $e^+e^-$ c.m. frame
for the two energetic tracks: $|\sum{\bf p}_t^*| =  
|{\bf p}_{t1}^* +{\bf p}_{t2}^*| < 0.1$~GeV/$c$.  

After the application of these selection criteria, 
$2.56 \times 10^7$ events remain.  They are dominated by
two-photon events of light-particle pairs: $\gamma \gamma \to e^+e^-$, 
$\mu^+\mu^-$ and $\pi^+\pi^-$.

We apply particle identification to the tracks in the remaining
events to select $\gamma \gamma \to K^+K^-$ events.
We suppress $e^+e^-$ events by requiring $E/p < 0.8$ 
for each track (where $E$ is the energy of the ECL cluster that 
matches the track of momentum $p$).  We require that
the TOF counter system gives useful time-of-flight information
for each track.
We suppress $\pi^+\pi^-$, $\mu^+\mu^-$ and $p\bar{p}$ events by
requiring that the TOF-, ACC-
and (for $K/p$ separation only) $dE/dx$-derived
particle identification
likelihood ratios ${\cal L}_K/({\cal L}_K+{\cal L}_\pi)$ 
and ${\cal L}_K/({\cal L}_K+{\cal L}_p)$ each exceed 0.8 for each 
track, and demanding $\Delta TOF_K > -0.6$~ns for each track,
where $\Delta TOF_K$ is the difference between the TOF-measured and calculated
times of flight of the charged track, assuming it to be a kaon.

An ambiguity in the determination of the collision time ($t_0$) 
introduces spurious events into our sample at low invariant mass
($M_{K^+K^-} < 1.55$~GeV/$c^2$). Our $t_0$ measurement in each event
initially has an ambiguity among times that differ by an integer multiple 
of the RF-beam-bucket spacing time ($t_s = 1.965$~ns), and we 
determine this offset in each event by requiring the consistent 
TOF identification of both tracks as a pair of known long-lived
charged particles. 
However, there is a kinematical region where the $t_0$ assignment 
has a two-fold solution,
differing by $\pm t_s$, that can lead to misidentification of 
$\pi^+\pi^-/\mu^+\mu^-/e^+e^-$ as $K^+K^-$ events or vice-versa.
To suppress the contamination from the lighter particle-pair events
where the wrong $t_0$ was chosen, we apply the following prescription.
Assuming that the two energetic tracks are pions, we calculate 
$\Delta TOF_{\pi}$ as well as
$\chi^2_{\pi} = (\Delta TOF_{\pi}/\sigma_t)^2$, where
$\sigma_t = 0.10$~ns is the nominal time resolution of the TOF, 
for each track using a collision time that is either nominal or shifted
forward or backward by $t_s$, then select
the case with the minimum $\chi^2_{\pi^+} + \chi^2_{\pi^-}$. 
We do the same, assuming both tracks to be kaons. If the optimal
collision time for the kaon hypothesis is one beam bucket interval 
earlier than the optimal time for the pion hypothesis, then we
reject the event if $\chi^2_{\pi^+} + \chi^2_{\pi^-} < 49$ or
$\Delta TOF_{\pi^+} + \Delta TOF_{\pi^-} > -0.73$~ns.
If the event survives this test, we also require
$\chi^2_{K^+} + \chi^2_{K^-}<25$ and that
the $dE/dx$ measurements in the CDC for each track be consistent with
a kaon.
  
Proton backgrounds
and the loss of signal due to $K\bar{K}$ misidentified as $p\bar{p}$
because of such $t_0$ misassignment are confirmed to be negligibly small 
by using a tighter $dE/dx$ cut and by investigating the
$\Delta TOF_K$ distribution, respectively.

Figure 1 shows the $\Delta TOF_K$ distribution of a track in an event
where the other track is positively identified as a kaon
and where both tracks are so identified. (The latter samples correspond to
the final signal candidates.)
The true kaons---the peak near $\Delta TOF_K = 0$~ns---are well separated 
from lighter particles (the peak below zero).
Clearly, the lighter-particle backgrounds are rejected while 
nearly all $K^+K^-$ events are retained by these cuts.
The largest remaining background contribution from particle misidentification 
comes from $\gamma \gamma \to f_2(1270) \to \pi^+\pi^-$, which
contaminates the signal sample up to 6\% in a narrow $K^+K^-$
invariant mass region around 1.60~GeV.

After the application of all the selection criteria,
63455 candidate events remain. Their invariant-mass
distribution is shown in Fig.~2. A peak at around 1.52~GeV comes from the
$f_2'(1525)$ resonance.  A decrease of the events below 1.45~GeV is mainly
due to an effect from the selection criteria for the $p_t$ cut of tracks
and the cut to avoid the $t_0$ determination ambiguity. A few bump structures
are seen at higher energies.

\begin{figure}

\resizebox{0.9\textwidth}{!}{%
  \includegraphics{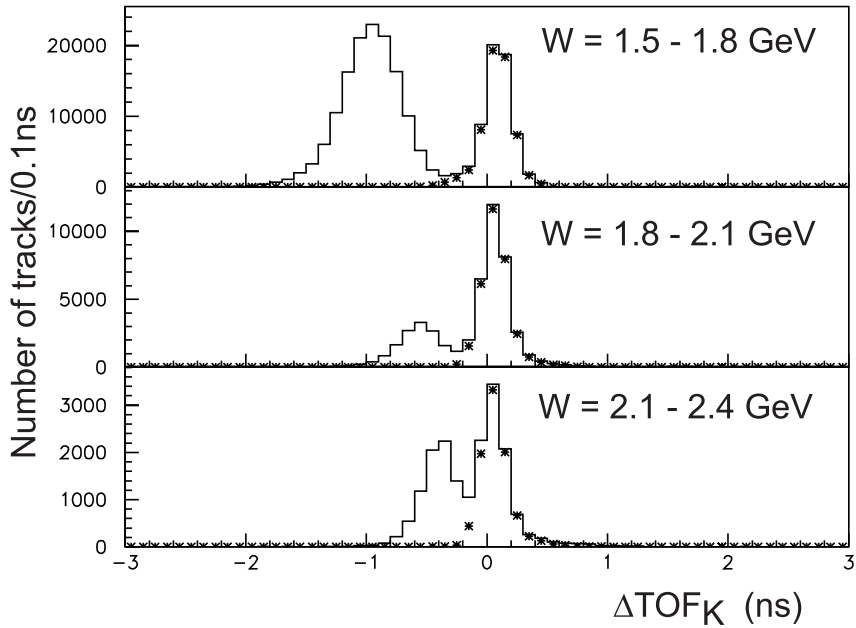}
}

\label{fig1}
\caption{
The $\Delta TOF_K$ distribution for each
track in the event where the other track
is identified as a kaon (histogram), and where both tracks
are so identified (asterisks),
in three different $W$ regions.}\end{figure}

\begin{figure}

\resizebox{0.9\textwidth}{!}{%
  \includegraphics{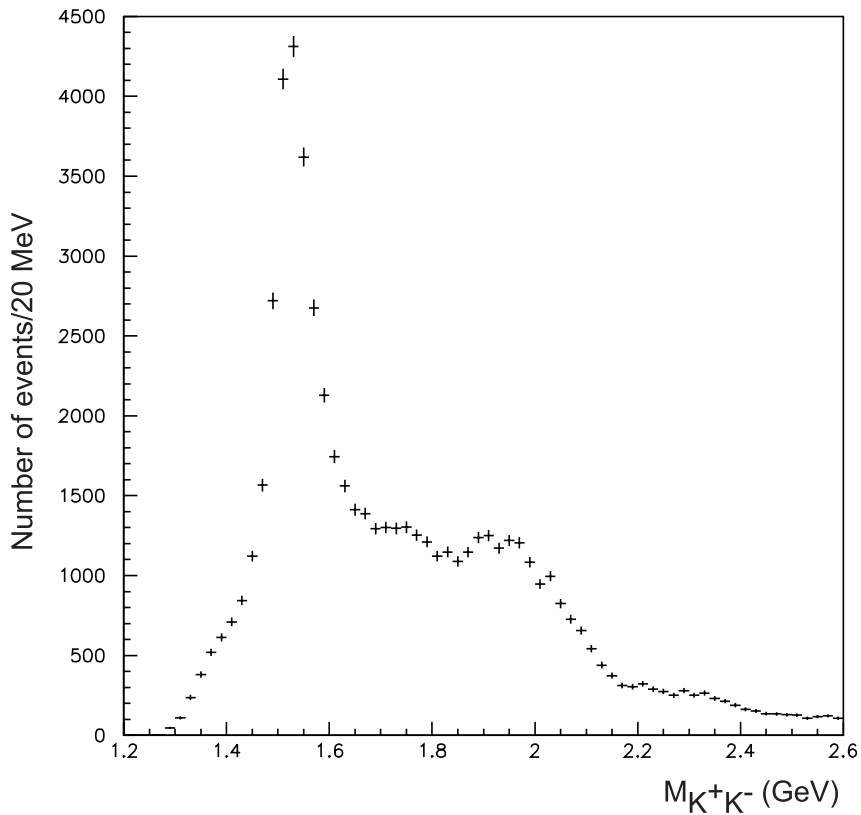}
}

\label{fig2}
\caption{
The invariant mass  distribution of 
the $K^+K^-$ candidates that satisfy all the
selection criteria.}\end{figure}

\section{Derivation of Cross Sections}
The cross section 
for $\gamma\gamma \to K^+K^-$ is derived from the present measurement. 
We restrict the polar-angle range of the final state kaons
in the $\gamma\gamma$ c.m. frame ($\theta^*$)  to be
within $|\cos \theta^*|<0.6$.  The differential cross sections
are given by the following formula:
\begin{eqnarray}
\frac{d\sigma}{d|\cos \theta^*|} = 
\frac{\Delta N(W,|\cos \theta^*|)~(1-f_{\rm BG}(W))}
{ \int{\cal L}dt~[L_{\gamma\gamma}(W) \Delta W]~\Delta |\cos \theta^*| 
~\eta(W,|\cos \theta^*|)},
\end{eqnarray}
where $\Delta N(W,|\cos \theta^*|)$ is the number of events
in a two-dimensional bin of the c.m. energy ($W$) and the cosine
of the polar angle of a meson in the c.m. system ($|\cos \theta^*|$)
with their widths $\Delta W$ and
$\Delta |\cos \theta^*|$, respectively, where we take the
absolute value of the cosine according to the symmetry of the
initial-state particles.
In the denominator, $\int{\cal L}dt$, $L_{\gamma\gamma}(W)$,  and 
$\eta(W,|\cos \theta^*|)$ are the integrated luminosity, 
the luminosity function and the event efficiency at the given 
kinematical point,
respectively.

Residual background due to particle misidentification is corrected
using an energy-dependent factor $f_{\rm BG}(W)$, that is derived from a study of
the $\Delta TOF_K$ distributions shown in Fig.~1.  This background
fraction is estimated to be less than 3\% except in the range 1.55 - 1.65~GeV
where it can be as large as 6\% due to $f_2(1270) \to \pi^+\pi^-$
contamination.
We neglect any potential $\theta^*$ dependence
of the correction factor since no prominent angular variation is anticipated
in the background that arises mainly from the non-Gaussian tails of
the TOF measurements.   

  The estimated contribution from other background processes, discussed
in section~5.3, is much smaller than the total systematic errors (shown in Table~2). 
We neglect the effect of such backgrounds in the 
derivation of the cross section.

We use the measured invariant mass of the $K^+K^-$ system as
the $\gamma\gamma$ c.m. energy $W$.
The fractional energy resolution $\Delta W/W \simeq 0.2\%$ is estimated from a
Monte Carlo (MC) simulation of the signal process (described below).
Since this is much smaller than
our energy bin size, we neglect smearing across bins
in the cross section derivation. 

Since we cannot determine
the true $\gamma\gamma$ axis in each event, we instead measure
$\theta^*$ from
the direction of the $e^+e^-$ beam axis in the $e^+e^-$ c.m. frame. 
The difference between this and the true
polar angle is confirmed to be small using the signal MC simulation,
corresponding to an r.m.s. deviation of about 0.015 in $\cos \theta^*$.

The efficiency is 
determined for each bin using the signal MC events 
for the $e^+e^- \to e^+e^-K^+K^-$ process that are generated by TREPS~\cite{treps}
and simulated within the Belle detector by a program based on GEANT3~\cite{geant}.
This efficiency is represented by a smooth 
function of ($W$, $|\cos \theta^*|$) in the analysis.

The trigger efficiency for $K^+K^-$ events is evaluated separately
via a study of detected two-photon $e^+e^-$ and $\mu^+\mu^-$ events that
satisfy two or more independent trigger conditions.
We parameterize the trigger efficiency as a function of
the averaged transverse momentum ($\bar{p_t}$) of the two tracks in an event. 
The transverse momentum difference is
at most 0.1~GeV/$c$ because of the selection criteria;
under these circumstances, the relation
$\bar{p_t} = \sqrt{(W/2)^2 - {m_K}^2}\sin \theta^*$ is quite accurate,
so the trigger efficiency is calculated by this formula
in each ($W$, $|\cos \theta^*|$) bin. The laboratory-angle dependence of the
trigger efficiency is small within the acceptance range, and we neglect
the effect.
 
  The typical value of the efficiency $\eta(W, |\cos \theta^*|)$ is 
13\% at large angles ($|\cos \theta^*| < 0.3$) in the high-invariant 
mass region ($W > 1.8$~GeV).   At most of the other points, it is
between 1\% and 10\%.  The large inefficiency seen in  $\eta(W, |\cos \theta^*|)$
is mainly attributed to detector acceptance effects. The efficiency to identify
kaon-pair events within the acceptance is typically 55\%, including losses from
kaon decays.

We calculate the luminosity function using a separate feature of TREPS~\cite{treps}. 
The effects of longitudinal photons are ignored therein.
We introduce the form factor effect for finite-$Q^2$ photons 
by multiplying the two-photon flux by the factor
$(1+Q_1^2/W^2)^{-2}(1+Q_2^2/W^2)^{-2}$
in the integrations over $Q_1^2$ and $Q_2^2$ for the
calculation of the luminosity functions.
(Here, $Q_1^2$ and $Q_2^2$ are the absolute values of the four-momentum 
transfer of each photon.)  The tight cut of 0.1~GeV/$c$ on 
$|\sum {\bf p}^*_t|$ in this analysis limits
the uncertainty from the choice of the form factor to less than 2\%.
The systematic error of the luminosity function is estimated to be
4\% by a study that compares the cross sections for the 
$e^+e^- \to e^+e^-\mu^+\mu^-$ process using this luminosity function
to those from a full $\alpha^4$-order QED calculation~\cite{berends}.
The luminosity function $L_{\gamma\gamma}(W)$ and the efficiency 
$\eta(W, |\cos \theta^*|)$ are consistently calculated for two-photon collisions
in the same $Q_1^2$ and $Q_2^2$ ranges, $Q_1^2$, $Q_2^2 <1$~GeV$^2$. 
The calculated $L_{\gamma\gamma}(W)$ is a smoothly decreasing function equal to
$5.9 \times 10^{-3}$~GeV$^{-1}$ at $W=1.4$~GeV and  $2.0 \times 
10^{-3}$~GeV$^{-1}$ at $W=2.4$~GeV.
  
The cross section values integrated over $|\cos \theta^*| < 0.6$ 
are derived by a binned maximum likelihood 
fit of the experimental $|\cos \theta^*|$
distribution with a bin width of 0.05 carried out
at each $W$ bin with a bin width of  0.02~GeV. 
We assume that the angular dependence of the differential cross 
section is a second-order 
polynomial of $\cos^2 \theta^*$ and integrate the fit
with $|\cos \theta^*|$ between 0 and 0.6.
Several data points at $W < 1.46$~GeV and
$|\cos \theta^*| > 0.50$ are removed from the fit,
where the efficiencies are extremely small.

The obtained cross sections for the process $\gamma\gamma \to K^+K^-$
integrated over the range $|\cos \theta^*|<0.6$ are 
summarized in Table~1 and depicted in Fig.~3.  The error bars in the figure
are statistical only. Systematic errors are tabulated in Table~2, and the
dominant sources are described in section~5.

  Figure 3 shows our result in comparison with the previous measurements.
The errors for our data are statistical only; we note that they are much
improved over those of ARGUS~\cite{cKK} and TPC/Two-Gamma~\cite{cKK2}.
The ARGUS cross sections are obtained for the full angular 
range ($|\cos \theta^*| \le 1 $) by fitting the angular distributions
to the sum of two spin-helicity components, $(J,\lambda)=(0,0)$ and 
$(2,2)$, and should therefore be somewhat larger---typically by
30\%---than those of our measurement and TPC/Two-Gamma.  (See
section~6 for the spin-helicity definition and decomposition.)

  Our differential cross sections in the $\gamma\gamma$ c.m. polar angle
are plotted in Fig.~4 for each energy bin of width $\Delta W = 0.04$~GeV.
Again, the displayed errors are statistical only.

\begin{table}
\caption{The cross section of the process $\gamma\gamma \to
K^+K^-$ in the polar angular region $|\cos \theta^*|<0.6$.
The first error is statistical and the second systematic.}
\begin{center}
\begin{tabular}{cc} 
\hline
$W$(GeV) & $\sigma$ ($|\cos \theta^*|<0.6$) (nb)\\
\hline
1.40 - 1.42 & $5.7 \pm	0.7  \pm 1.2$  \\
1.42 - 1.44 & $5.4 \pm	0.7  \pm 1.0$   \\ 
1.44 - 1.46 & $5.4 \pm	0.5  \pm 0.8$   \\
1.46 - 1.48 & $7.1 \pm	0.3  \pm 1.0$   \\
1.48 - 1.50 & $10.0 \pm 0.2  \pm 1.4$ \\
1.50 - 1.52 & $14.4 \pm 0.3  \pm 2.0$  \\
1.52 - 1.54 & $14.2 \pm 0.3  \pm 1.9$   \\
1.54 - 1.56 & $11.4 \pm 0.2  \pm 1.6$   \\
1.56 - 1.58 & $7.9 \pm 0.2  \pm 1.0$    \\
1.58 - 1.60 & $5.98 \pm 0.15  \pm 0.77$ \\
1.60 - 1.62 & $4.81 \pm 0.13  \pm 0.64$ \\
1.62 - 1.64 & $4.22 \pm 0.12  \pm 0.56$ \\
1.64 - 1.66 & $3.70 \pm 0.11  \pm 0.48$  \\
1.66 - 1.68 & $3.55 \pm 0.11 \pm 0.44$  \\
1.68 - 1.70 & $3.31 \pm 0.11 \pm 0.41$  \\ 
1.70 - 1.72 & $3.27 \pm 0.10 \pm 0.41$  \\
1.72 - 1.74 & $3.29 \pm 0.11 \pm 0.41$  \\
1.74 - 1.76 & $3.23 \pm 0.10 \pm 0.40$ \\
1.76 - 1.78 & $3.00 \pm 0.10 \pm 0.37$  \\ 
1.78 - 1.80 & $2.94 \pm 0.10 \pm 0.36$  \\
1.80 - 1.82 & $2.62 \pm 0.09 \pm 0.32$  \\ 
1.82 - 1.84 & $2.71 \pm 0.09 \pm 0.32$  \\ 
1.84 - 1.86 & $2.56 \pm 0.09 \pm 0.30$  \\
1.86 - 1.88 & $2.64 \pm 0.09 \pm 0.31$  \\ 
1.88 - 1.90 & $2.92 \pm 0.09 \pm 0.34$  \\ 
1.90 - 1.92 & $2.95 \pm 0.09 \pm      0.35$ \\
1.92 - 1.94 & $2.63 \pm 0.08 \pm      0.31$ \\ 
1.94 - 1.96 & $2.84 \pm 0.09 \pm 	0.33$ \\
 1.96 - 1.98 & $2.92 \pm 0.10 \pm 	0.34$ \\
1.98 - 2.00 & $2.59 \pm 0.09 \pm 	0.31$ \\
 2.00 - 2.02 & $2.26 \pm 0.08 \pm 	0.26$ \\
 2.02 - 2.04 & $2.36 \pm 0.08 \pm 	0.27$ \\
 2.04 - 2.06 & $1.96 \pm 0.08 \pm 	0.23$ \\
 2.06 - 2.08 & $1.68 \pm 0.07 \pm 	0.20$ \\
 2.08 - 2.10 & $1.57 \pm 0.07 \pm 	0.18$ \\
2.10 - 2.12 & $1.34 \pm 0.06 \pm 	0.17$ \\
 2.12 - 2.14 & $1.15 \pm 0.06 \pm 	0.14$ \\
 2.14 - 2.16 & $0.96 \pm 0.06 \pm 	0.12$ \\
 2.16 - 2.18 & $0.89 \pm 0.06 \pm 	0.11$ \\
 2.18 - 2.20 & $0.89 \pm 0.06 \pm 	0.11$ \\ 
 2.20 - 2.22 & $0.97 \pm 0.06 \pm 	0.12$ \\
 2.22 - 2.24 & $0.90 \pm 0.06 \pm 	0.11$ \\
 2.24 - 2.26 & $0.83 \pm 0.06 \pm 	0.10$ \\
 2.26 - 2.28 & $0.84 \pm 0.06 \pm 	0.10$ \\ 
 2.28 - 2.30 & $0.96 \pm 0.06 \pm 	0.12$ \\
 2.30 - 2.32 & $0.80 \pm 0.05 \pm 	0.10$ \\ 
 2.32 - 2.34 & $0.93 \pm 0.06 \pm 	0.12$ \\ 
 2.34 - 2.36 & $0.76 \pm 0.06 \pm 	0.09$ \\
 2.36 - 2.38 & $0.79 \pm 0.06 \pm 	0.10$ \\ 
 2.38 - 2.40 & $0.68 \pm 0.06 \pm 	0.08$ \\ 
\hline
\end{tabular}
\end{center}
\end{table}

\begin{table}

\caption{ The systematic errors for the
cross section $\gamma\gamma \to
K^+K^-$ in the polar angular region $|\cos \theta^*|<0.6$.
Some systematic errors are $W$-dependent;
these are shown as ranges.}
\begin{center}
\begin{tabular}{cc} 
\hline
        & Systematic\\ 
Source       & error\\    
\hline
Trigger efficiency & 4 - 6\% \\
Tracking efficiency & 4\% \\
$E/p$ cut & 4 - 6\% \\
TOF efficiency & 3\% \\
Kaon identification efficiency by ACC & 0 - 5\% \\
$t_0$ ambiguity & 0 - 7\% \\
Acceptance calculation & 3 - 15\% \\
Particle misidentification backgrounds & 1 - 2\% \\
Non-exclusive backgrounds & 4 - 6\% \\
Integrated luminosity & 1\% \\ 
Luminosity function and form-factor effect& 4\% \\
\hline
Total & 11 - 20\% \\
\hline
\end{tabular}
\end{center}
\end{table}

\begin{figure}
\resizebox{0.9\textwidth}{!}{%
  \includegraphics{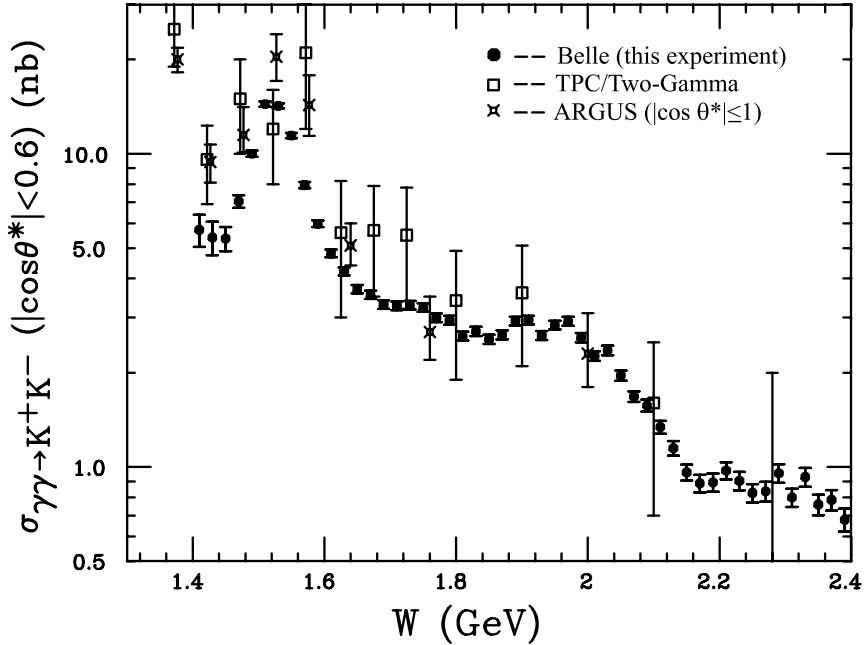}
}
\label{fig3}
\caption{The cross section for $\gamma\gamma \to
K^+K^-$ in the polar angular region $|\cos \theta^*|<0.6$ 
obtained in this experiment
compared with results from previous experiments~[1,2].
The error bars are statistical only.}
\end{figure}

\begin{figure*}
\begin{center}
\resizebox{0.95\textwidth}{!}{%
 \includegraphics{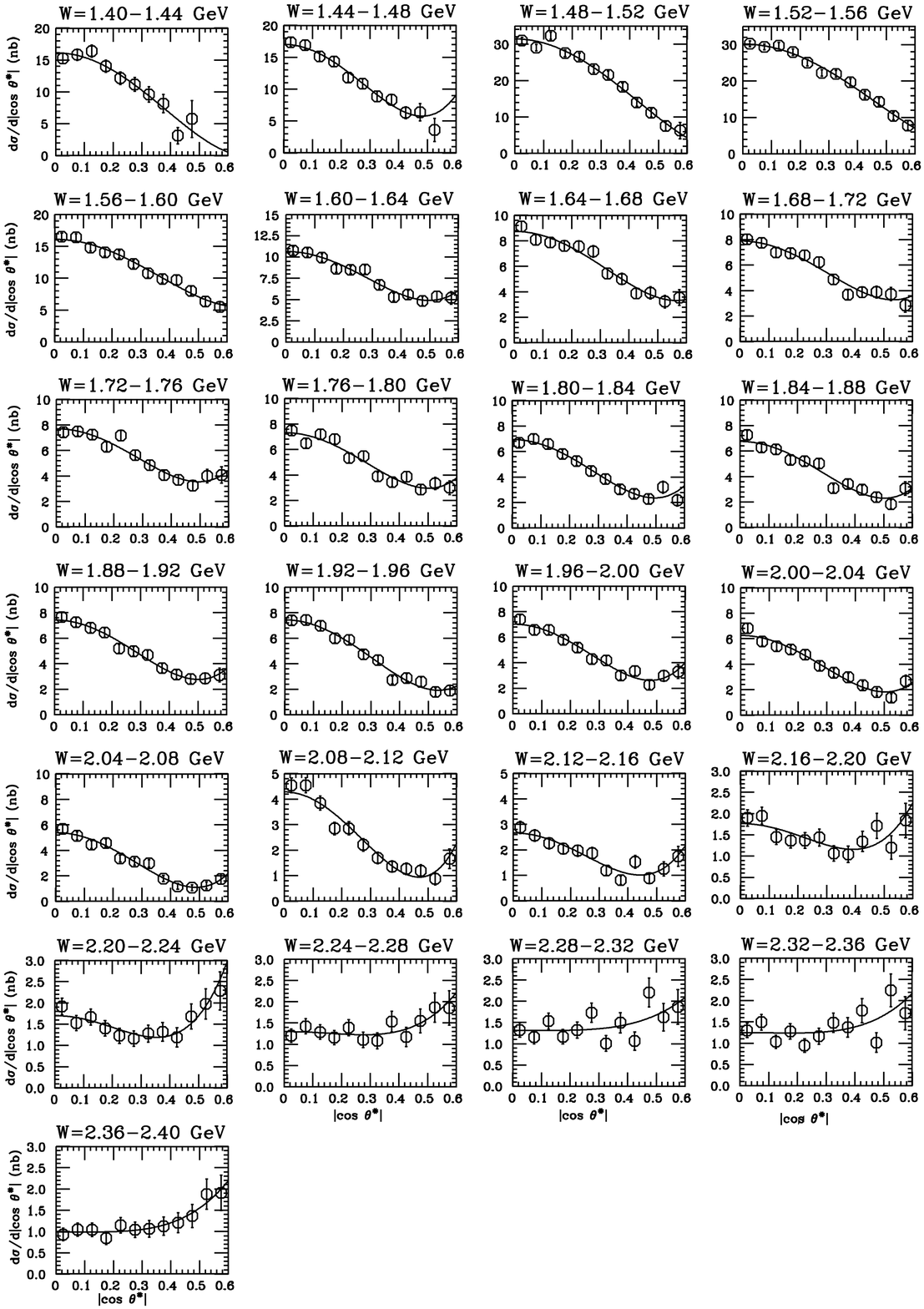}
}

\label{fig4}
\caption{The differential 
cross section $d\sigma/d|\cos \theta^*|$
for $\gamma\gamma \to K^+K^-$
obtained in this experiment. The solid curves
are the fits described in section~6.2. The error bars
are statistical only.}
\end{center}
\end{figure*}

\section{Major sources of systematic error}
\subsection{Trigger efficiency}
  We determine the trigger efficiency for the $K^+K^-$ final 
samples that would pass all the selection criteria described in
section~3.  We first estimate the efficiency of the two-track trigger,
which requires two or more TOF/TSC hits as well as one or more
CsI clusters.  The efficiency is determined
experimentally as a function of transverse momentum between 0.4 and 1.2~GeV/$c$, using
the redundancy of triggers for two-photon
$e^+e^-$ and $\mu^+\mu^-$ events.  The
ratio of the yield from this particular trigger
to that from all the trigger sources ---typically 0.88--- is
used to determine the overall trigger efficiency.
We confirm that there is no notable explicit dependence of the trigger efficiency
on the polar angle in the laboratory system. The results on the 
trigger efficiency are compared with
those from our trigger simulation program, which is applied immediately after the full
detector simulation.  
The simulation reproduces a quantitative nature of the $p_t$ dependence of 
the trigger efficiency.  Although the trigger efficiency value from the 
simulation is not used for the cross section calculations, the
difference between experimental and MC values is taken into account in the 
systematic error. A small difference between the 
trigger efficiency for leptons and hadrons estimated from the experimental and 
simulation studies is applied as a correction and included to the 
systematic error as well.
The trigger efficiency thus determined
is $(82 \pm 5)\%$ at $\bar{p_t}=0.55$~GeV/$c$ and $(92 \pm 4)\%$ at
$\bar{p_t}=1.0$~GeV/$c$. 

  The samples after application of the event selection criteria
{\it before particle identification} are used for a confirmation of
the trigger efficiency. We have compared the experimental
yields of these samples in the $W$ and laboratory polar angle
distributions to the expectations from the MC calculation.  Those samples
are dominated by $e^+e^- \to
e^+e^-e^+e^-$, $e^+e^-\mu^+\mu^-$ and $e^+e^-\pi^+\pi^-$ processes 
whose cross sections are well known~\cite{berends,pipiexp,yabuki}. 
The experimental and MC yields are consistent within the above error of 
the trigger efficiency.

\subsection{Kaon-identification efficiency}
We have checked the kaon-identification efficiency in this analysis
based on the real data. The efficiency to identify and select kaon-pair events
within the detector acceptance is typically 55\%. The 45\% loss is partially
due to kaon decays, which account for 20-30\% depending on the two kaon invariant mass.
An additional reduction of $\sim 25$\% comes from the kaon identification
criteria for the TOF, ACC and $E/p$ requirements.

The efficiency of the TOF counters
giving a useful TOF measurement
for a track, which is required in the kaon selection,
is obtained using the events in which the other track is tagged as a kaon.
The efficiency thus obtained in the real data 
is around 92\% in the $0.5 < p <1.0$~GeV/$c$ region 
where the signal events dominate, comparable to the MC simulation's
efficiency of 94\% at around 1~GeV/$c$. The 
effect of a {\it trigger} efficiency loss due to kaon decays is small
in this comparison.\footnote{
The effect of kaon decays in flight is taken into account in the simulation, and it appears
as a loss of the TOF efficiency (and thus, the loss of the $K^+K^-$ signals)
in both experimental and simulated data.
However, a few of the decays in the experimental data also induce a loss 
at the trigger stage; direct comparison is impossible
in the low-energy region.}
We take the difference of the TOF efficiency between the 
experimental and MC data, 3\% for the two tracks, as a systematic error.

  The efficiency of kaon-pair identification in the offline selection is examined using
the redundancy of two kaons in a signal $K^+K^-$ event. The $\Delta TOF_K$ 
distribution for a track is investigated for 
events where the other track is identified as a kaon as shown in Fig.~1.
We find that charged tracks whose $\Delta TOF_K$ is consistent with a kaon are
identified as a kaon with an efficiency higher than 95\%; 
this value is consistent with the MC expectation within the systematic
error of 5\% or less in all $W$ regions.

 The $E/p$ cut gives an inefficiency in kaon 
selection as a nuclear interaction of a negative kaon
in the ECL sometimes gives a large energy deposition.  We
compared the fractions of kaons with $E/p > 0.8$ to those with any
$E/p$ between the MC
events and real $K^+K^-$ samples; they are consistent. 
A small possible discrepancy ($\sim 6\%$ at $W<1.8$~GeV
and $\sim 4\%$ at $W>1.8$~GeV) is accounted for in the systematic error.

\subsection{Backgrounds from particle misidentification and non-\\
exclusive events}
The correction factor for the
background contamination from particle misidentification
is evaluated from the study of the experimental data,
as described in section~4.  Another study using the MC expectations from
the $e^+e^-e^+e^-$, $e^+e^-\mu^+\mu^-$~\cite{berends} and $e^+e^-\pi^+\pi^-$
processes~\cite{pipiexp,yabuki}, which dominate this kind of background, 
is also consistent with this evaluation. We assign a 
systematic error
of 1-2\% from this source, depending on $W$. This background
source has no effect on the significance of the resonant structures 
that are observed in the $K^+K^-$ spectrum
and discussed in this paper.

 We also estimate the background contamination from events where
additional particles accompany the two detected tracks---so-called
non-exclusive backgrounds. Such events are expected to give a
larger $|\sum {\bf p}^*_t|$ than the signal events.
Figure~5 shows the distribution of $|\sum {\bf p}^*_t|$ for the $K^+K^-$
signal candidates, with the simulated distribution scaled to match the
data yield in the first two bins.
The data and signal MC show generally good
agreement over the entire accepted range of $|\sum {\bf p}^*_t|$,
and we conclude that the backgrounds from non-exclusive processes
are small. The small difference between the experimental and MC 
distributions seen at high $|\sum {\bf p}^*_t|$ may be attributed
to the non-exclusive backgrounds (or any other process that does not have
an enhancement at zero-$p_t$ balance).

\begin{figure}
\resizebox{0.87\textwidth}{!}{%
  \includegraphics{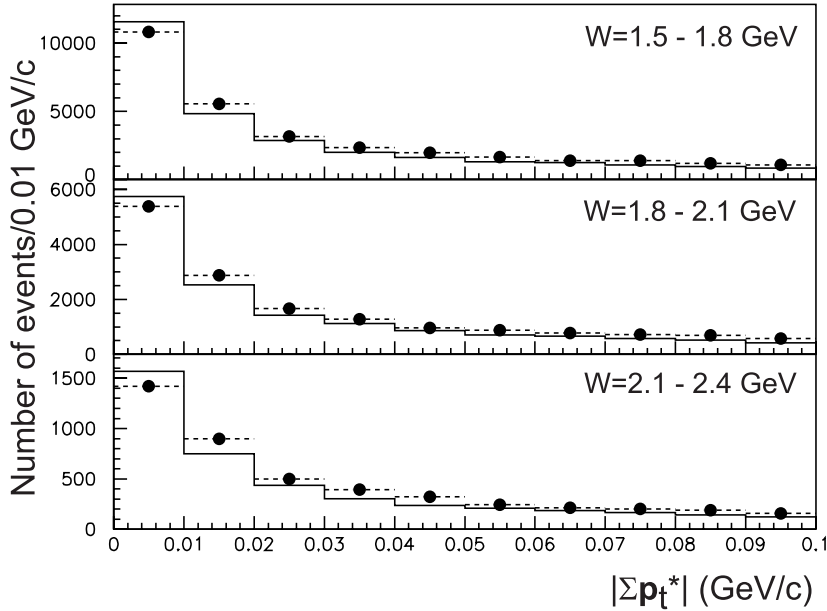}
}

\label{fig5}
\caption{ The transverse momentum balance ($|\sum {\bf p}^*_t|$)
distribution for the signal process from the
real data (closed circles) and signal MC events (histogram)
for events in three $W$ regions.
The MC distributions are scaled to match the data yield in
the first two bins. }
\end{figure}

  We find that the contamination of such background processes is less than 6\% at
any $W$, based on the expectation that the background contribution vanishes 
at $|\sum {\bf p}^*_t|=0$ and the assumption that 
the observed surplus over MC expectation 
in the rightmost bin of Fig.~5 is purely from 
the contribution from the non-exclusive background.
In the experimental data, a larger excess over the MC expectation is seen at   
$|\sum {\bf p}^*_t|$ around 0.01 - 0.04 GeV/$c$.
This excess cannot be attributed to the non-exclusive backgrounds,
and is considered to be due to an unmodelled broadening of the
signal-process distribution.\footnote{This broadening may be due to
interferences or higher-order processes in the relevant 
photon-emission diagrams, as well as a mismatch in the real and simulated
resolution of the detector components.}

  It is difficult to determine the magnitude of these backgrounds, so
we neglect their effect in the derivation of cross sections and instead account for them
in the systematic error.

\section{Phenomenological analyses}
\subsection{The c.m.-energy dependence}
 The obtained $\gamma\gamma$ c.m.-energy dependence of the
experimental yield (Fig.~2) and 
the cross section (Fig.~3) clearly show
four resonant structures, in the vicinities of 
1.52~GeV, 1.7 - 1.8~GeV, 1.9 - 2.1~GeV, and 2.2 - 2.4~GeV.
The peak near 1.52~GeV
is from the $f'_2(1525)$ resonance, which is considered to be
a ground-state tensor meson dominated by an $s\bar{s}$ component.
The contribution of this resonance in this reaction is well known from previous 
measurements~\cite{cKK,cKK2}.
In contrast, this measurement constitutes the first observation of the other
three structures.

 The $W$ dependence of the cross section is fitted to a sum of contributions
from resonances and a  continuum component. The components adopted in the 
fit are: (A) the $f'_2(1525)$ resonance, 
(B) a resonance around 1.75~GeV, (C) a resonance around 2.0~GeV,
(D) a resonance around 2.3~GeV, 
(E) contributions from the continuum component and the tails of two low-mass
resonances, $f_2(1270)$ and $a_2(1320)$.  For (A)--(D), we
use an amplitude with a relativistic Breit-Wigner form:
\begin{eqnarray}
{\cal A}_R = \frac{\sqrt{S \Gamma}}{W^2 - M^2 - iM\Gamma},
\end{eqnarray}
where $S$, $M$ and $\Gamma$
are the size parameter, mass and total width of the resonance, respectively. 
The component (E) is expressed by an amplitude with a
$W$-dependent form and three free parameters $a$, $b$, and $W_0$ as 
\begin{eqnarray}
{\cal A}_E = a (\frac{W-W_0}{1.4~{\rm GeV}-W_0})^{-b},
\end{eqnarray}
where we combine the effects from the tails of the low-mass resonances and
the smooth continuum since we cannot decompose
these effects in our present measurement. The interference
phases between (A) and (E) and between (C) and (E) are
treated as free parameters. (B) and (D)
are considered incoherent since \ their \ peaks
\ are \ relatively \ small and \ it is \ difficult to \ determine
\ the \ relative \ phase.\footnote{Moreover, it is 
noted that the interference term from two distinct resonance amplitudes
vanishes in the present process in case the resonances are 
in different helicity states.}
For \ the \ $f'_2(1525)$ \ resonance, \ the \ $W$-dependent \ total \ width \ is 
$\Gamma(W) = \Gamma_M(p^*/p_0^*)^5 (M/W) (D_2(p^*r)/D_2(p_0^*r))$,
where $M$ and $\Gamma_M$
are the nominal mass and  total width, respectively,
while $p^*$ and $p_0^*$ are the momenta 
of the final-state kaon in the rest
frame of the resonance with invariant mass $W$ and $M$, respectively.
The function $D_2(z) \sim (9+3z^2+z^4)^{-1}$~\cite{decayf}
represents a centrifugal barrier factor; we use
an effective interaction radius of $r=1$~fm.
We use $W$-independent total widths for the other three resonances.
We denote the $\Gamma_M$ for the $f_2'(1525)$ simply by $\Gamma$ in the 
remainder of this paper.

The result of the fit is summarized in Table 3. 
The curves in Fig.~6 show the fit and the contribution from each component 
superimposed on the experimental data.
The goodness of fit is $\chi^2/{\rm ndf}=50.8/33$. We use only
statistical errors for this calculation and its minimization. 
We determine the systematic errors of the fitted parameters by a study
of their changes for four cases of linear deformations of 
the measured cross section as a function of $W$.\footnote{
In the first two cases, we shift the cross sections at the first and last $W$ bins 
by $\pm 1.5\sigma$ of the systematic errors in opposite directions and make 
a linear deformation of the cross sections at the other intermediate bins.
In the second two cases, we shift the cross sections 
by $\pm 1.5\sigma$ in the same direction at the two end bins and in the opposite 
direction at the center bin. We treat the largest
observed deviation of each fit parameter from its original value as the 
systematic error.}

\begin{table*}
\caption{Resonance parameters and other variables 
obtained from the fit of the $W$ dependence of the cross section. 
The first error is statistical and the second systematic.}
\begin{center}
\begin{tabular}{cccccc} 
\hline
\multicolumn{2}{c}{Resonance} & Mass & Total width & $S$ & Signi- \\
\multicolumn{2}{c}{components} &   (MeV/$c^2$)      & (MeV)  &  (nb~GeV$^3$) & ficance\\    
\hline
(A)&($f_2'(1525)$) & $1518 \pm 1 \pm 3$ & $82 \pm 2 \pm 3$ & $1.22 \pm 0.11 \pm 0.25$ & $>25\sigma$\\
(B)&(1.75~GeV) & $1737 \pm 5 \pm 7$ & $151 \pm 22 \pm 24$ & $0.45 \pm 0.09 \pm 0.10$ & $5.5\sigma$ \\
(C)&(2.0~GeV) & $1980 \pm 2 \pm 14$ & $297 \pm 12 \pm 6$ & $2.63 \pm 0.11 \pm 0.57$ & $>9\sigma$ \\
(D)&(2.3~GeV) & $2327 \pm 9 \pm 6$ & $275 \pm 36 \pm 20$ & $0.94 \pm 0.13 \pm 0.28$ & $5.3\sigma$ \\
\hline
\hline
\multicolumn{2}{c}{(E)} & 
\multicolumn{4}{c}{$a=2.12 \pm 0.17 \pm 0.54$~nb$^{1/2}$,}\\ 

\multicolumn{2}{c}{ } & 
\multicolumn{4}{c}{
$W_0 = 0.04 \pm 0.02 \pm 0.33$~GeV,
$b = 7.9 \pm 1.0 \pm 2.0$ }\\
\hline
\multicolumn{2}{c}{Interference} &
\multicolumn{4}{c}{${\rm (A)}-{\rm (E)}: -1.27 \pm 0.07 \pm 0.12$~rad}\\
\multicolumn{2}{c}{phases } &
\multicolumn{4}{c}{${\rm (C)}-{\rm (E)}: +2.57 \pm 0.14 \pm 0.38$~rad}\\
\hline
\end{tabular}
\end{center}
\end{table*}

The small confidence level of the  $\chi^2/{\rm ndf}$ value is
attributed to the inability of our fit function to reproduce faithfully
the true behavior of the cross section rather than any
systematic-error effects; nevertheless, we expect that we can obtain
meaningful resonance parameters by this parametrization.
We note that there could be additional systematic shifts in the
parameters for the particular resonant structure at 2.3~GeV due to the boundary
of our available $W$ range, $W < 2.4$~GeV.

The statistical significances for the resonant structures are also
given in the last column of Table 3, They are derived from the square
root of the goodness of fit difference between the fits with and without
the corresponding resonance components for the structures at 
1.75 GeV and 2.3 GeV.  Each of the other two structures dominates the
cross section in its energy region, and this technique is not suitable
to estimate the significance.  We show, instead, the significances for the 
excess of the core part of the peak that appears above the averaged level of 
whole the apparent peak structure, regarding the deviation from the level
as the corresponding lower limit.

The component (E) gives a significant contribution only at $W < 1.7$~GeV
in the above fit.  An alternative form for (E) with two summed amplitudes
having distinct tail decay parameters did not yield an improved goodness of fit nor a
substantial change in the overall shape of the (E) component and the values of the
other resonance parameters.
We conclude that the cross section at energies above 1.7~GeV is dominated
by the contribution from the four resonances that appear in our fit.

  We note that the size parameter $S$ is proportional to the product
$\Gamma_{\gamma\gamma}(R){\cal B}(R \to K^+K^-)$ 
of the resonance and depends on its spin $J$ and helicity $\lambda$ along
the $\gamma\gamma$ axis as:

\begin{eqnarray}
S = 8\pi(2J+1)F~\Gamma_{\gamma\gamma}(R) {\cal B}(R \to K^+K^-),\\
F=(2J+1)\sum_{\lambda=0,2}f_\lambda \int_0^{0.6}[d^J_{\lambda,0}(z)]^2dz.
\end{eqnarray}
$F$ corresponds to the fraction of the resonance component's cross section
integrated over $|\cos \theta^*| <0.6$ to the total,
while $f_\lambda$ is
the fraction of the resonance production rate in the $\lambda$
helicity state (with $f_0 + f_2 = 1$).  The integrands in Eq.~(5) are the square of the 
factors of the Wigner $D$-functions. For a $J=0$ meson,
only $\lambda=0$ is allowed so that $F=0.6$. We discuss the extraction
of $\Gamma_{\gamma\gamma}(R){\cal B}(R \to K^+K^-)$ from this analysis
in section~7.

\begin{figure}
\resizebox{0.9\textwidth}{!}{%
  \includegraphics{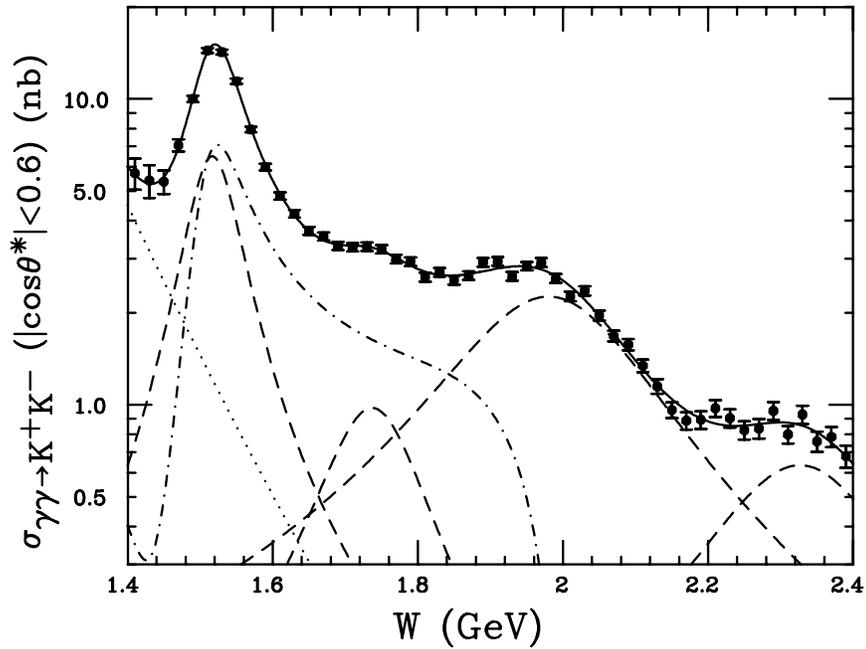}
}

\label{fig3b}
\caption{The measured cross section of the process $\gamma\gamma \to
K^+K^-$ in the polar angular region $|\cos \theta^*|<0.6$, 
compared with the fit described in section~6.1. 
The error bars are statistical only.
The solid curve indicates the best fit.
The dotted curve and four dashed curves
indicate the contributions from component (E) and from the 
four resonance components, respectively, without interference.
The dot-dashed curve is the contribution of 
the interference term.}
\end{figure}

\subsection{Angular dependence}
 The angular dependence of the differential cross section has
an enhancement at large angles (near $|\cos \theta^*|=0$)
for the lower values of $W$, but has a tendency
of a forward enhancement at the higher $W$ values.
The dependence
is almost flat in the vicinity of 2.3~GeV.
We first plot the ratio of cross sections over the restricted range
$|\cos \theta^*|<0.3$ and the measured range ($|\cos \theta^*|<0.6$):

\begin{eqnarray}
R=\frac{\int_{0}^{0.3} \frac{d\sigma}{d|\cos \theta^*|} d|\cos \theta^*|}
{\int_{0}^{0.6} \frac{d\sigma}{d|\cos \theta^*|} d|\cos \theta^*|}
\end{eqnarray}
to draw out this gross behavior of the angular distribution.
The $W$ dependence of this ratio is plotted in Fig.~7 
for $W$ points above 1.46~GeV,  where we
have complete angular data over the range $|\cos \theta^*|<0.6$. 
It has a large value at around 1.5 GeV and 2.0 GeV, respectively. 
These positions correspond to the two larger resonant structures designated
(A) and (C) in section 6.1.

 We make a model-independent angular analysis for
the angular distribution of the $K^+K^-$ events. 
The angular distribution is fitted by the sum of the
partial-wave contributions, independently at each $W$ bin.
For the total angular
momentum ($J$) of the $\gamma\gamma$ system, we assume
that the partial waves with $J \ge 4$ are negligible. 
The odd-$J$ waves are forbidden by the symmetric nature 
of the initial state and parity conservation. 
In the $\gamma\gamma$ c.m. system, the differential cross 
section can be written as

\begin{eqnarray}
\frac{d \sigma}{d \Omega} = |{\cal H}(0,0) +
{\cal H}(2,0)|^2 + |{\cal H}(2,2)|^2
\end{eqnarray}
where ${\cal H}(J,\lambda)$ is the partial-wave amplitude for the 
(spin, helicity)=($J$, $\lambda$) state, and $\Omega$ is the solid angle. 

Equation~(7) includes four real parameters---the sizes of the
three partial-wave amplitudes and one relative phase between
the two $\lambda=0$ waves---while only three are 
determined independently by this measurement.
(This formula is a quadratic
function of $\cos^2\theta^*$.) Thus, we choose
an alternative parameterization as follows.
With $\sigma_{J\lambda} = \int|{\cal H}(J,\lambda)|^2 d\Omega$
as the positive definite squared partial wave amplitude over the whole
solid angle, we parameterize the differential cross section in terms of three
cross section-like parameters $\sigma_a$, $\sigma_b$, and $\sigma_c$
and the factors of the Wigner $D$-functions as~\cite{yabuki}
\begin{eqnarray}
\frac{d\sigma}{d|\cos \theta^*|} = \sigma_a + 5\sigma_b\{d^2_{0,0}(\cos \theta^*)\}^2
 + 5\sigma_c\{d^2_{2,0}(\cos \theta^*)\}^2.
\end{eqnarray}
The coefficients $\sigma_a$, $\sigma_b$ and $\sigma_c$ coincide with $\sigma_{00}$,
$\sigma_{20}$ and $\sigma_{22}$, respectively, if there is no
interference between the two $\lambda=0$ components.
When we account for this interference,
the correspondence is
modified~\cite{yabuki}: 

\begin{eqnarray}
\sigma_a &=& \sigma_{00} +  s_I \\
\sigma_b &=& \sigma_{20} + \frac{1}{5}s_I\\
\sigma_c &=& \sigma_{22} - \frac{6}{5}s_I\\
s_I &=& \sqrt{5\sigma_{00}\sigma_{20}}\cos \psi,
\end{eqnarray}
where $\psi$ is the relative phase between the interfering amplitudes
${\cal H}(0,0)$ and ${\cal H}(2,0)$ that we cannot determine
independently.  We note that 
$\sigma_a$, $\sigma_b$ and $\sigma_c$ can be negative in the presence
of this interference.

  The fit values for $\sigma_a$,
$\sigma_b$ and $\sigma_c$ in each $W$ bin  are 
summarized in Fig.~8. The corresponding fit curves 
are shown in Fig.~4, which indicates a good parametrization
of the measured magnitude and polar angular dependence.

In the fits, we constrain
the best-fit values of $\sigma_a$, $\sigma_b$, and $\sigma_c$ 
to give a non-negative differential cross section
throughout the full angular range, $|\cos \theta^*| \leq 1$.
Only the fit to the lowest-energy bin, at $W= 1.40$ - 1.44~GeV,
is affected by this constraint
($d\sigma/d|\cos \theta^*|=0$ near $|\cos \theta^*|=0.67$).

\begin{figure}
\resizebox{0.9\textwidth}{!}{%
  \includegraphics{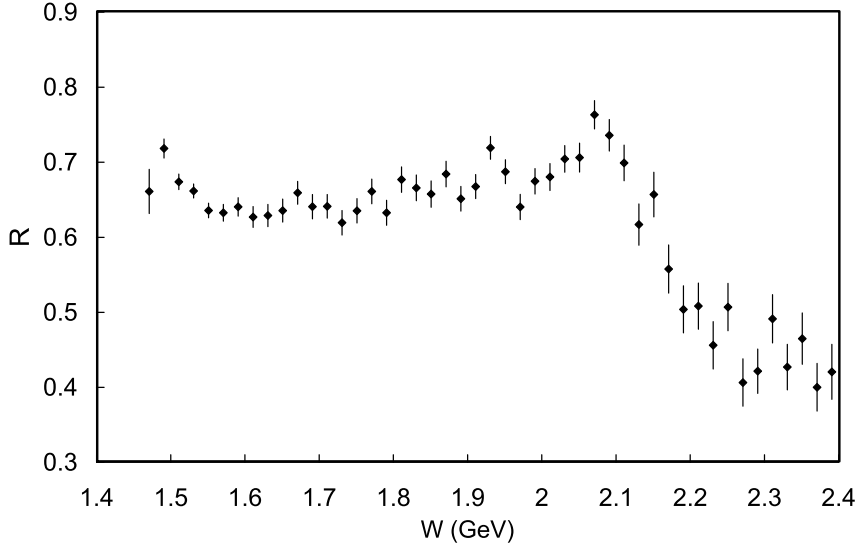}
}

\label{fig6}
\caption{The $W$ dependence of $R$, the ratio of
the cross sections in the angular ranges $|\cos \theta^*|<0.3$
and $|\cos \theta^*|<0.6$.}
\end{figure}

\begin{figure}
\resizebox{0.9\textwidth}{!}{%
  \includegraphics{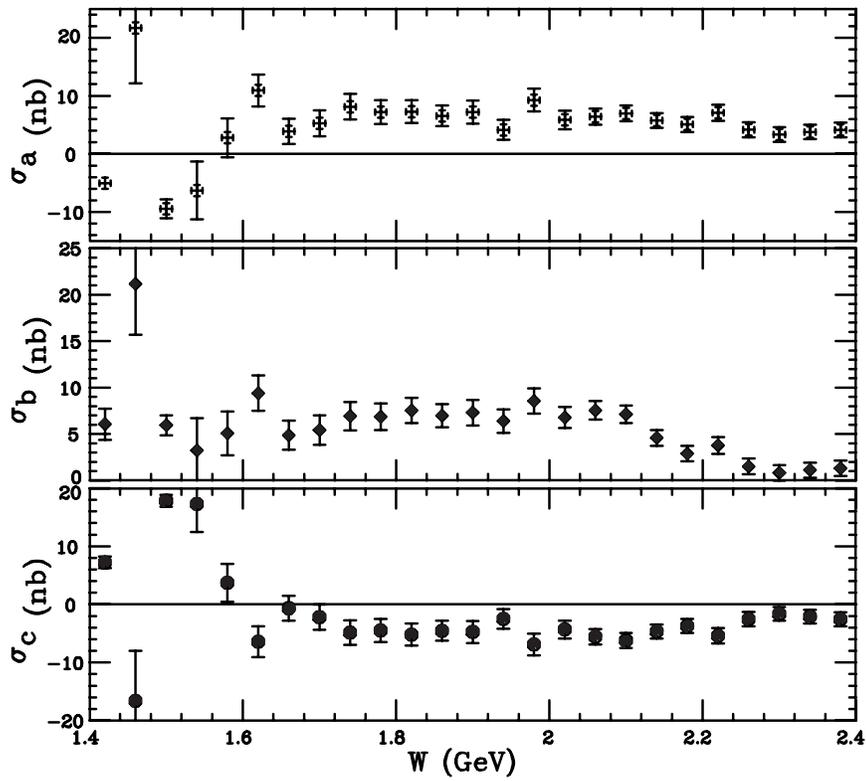}
}

\label{fig7}
\caption{The solution for the partial-wave coefficients $\sigma_a$ (top),
$\sigma_b$ (middle), and $\sigma_c$ (bottom).}
\end{figure}

\section{Discussion}
\subsection{{\boldmath $f_2$'(1525)}}

The resonance parameters obtained by the present
measurement are summarized and compared with those from previous experiments
in Table~4.  

Our best fit mass and total width of $f'_2(1525)$ are in agreement
with those from previous experiments~\cite{pdg}. 
In the range 1.5 - 1.6~GeV, the parameter $\sigma_c$ has a large peak. 
This feature is consistent with the previous determination that the
$f'_2(1525)$ is a spin-2 meson. We confirm that its
production in two-photon collisions is dominated by the $\lambda=2$
helicity component.

  From each size parameter $S$, we
extract $\Gamma_{\gamma\gamma}(R){\cal B}(R \to K^+K^-)$ for
the corresponding resonance $R$ of assigned spin and helicity
using Eqs.~(4) and (5). These results are also shown in Table~4.
Their systematic errors
are simply scaled from those in $S$; they do not
incorporate any effects from the assumptions in
resonance formulae, background shapes and interference effects.

When we assume a pure $(J,\lambda)=(2,2)$ state for $f'_2(1525)$, we obtain 
$\Gamma_{\gamma\gamma}(f'_2){\cal B}(f'_2 \\
\to K^+K^-)= 28.2 \pm
2.4\ {\rm (stat.)} \pm 5.8\ \rm{(syst.)} $~eV.
This result is slightly smaller than the world average~\cite{pdg} 
(where isospin invariance is assumed) but is 
still larger than the ARGUS result
where the interference effect is taken into account~\cite{cKK}.

We note that our systematic error does not account fully for the ambiguity in
the interference effect with the low-lying resonances
$f_2(1270)$ and $a_2(1320)$,
since this effect cannot be clarified by the present data  
alone.  We have tried an alternative fit where the mass, width and size parameters of 
the $f_2(1270)$ and $a_2(1320)$ are fixed to their accepted values~\cite{pdg}
and they are assumed to interfere with a zero relative phase;  
the quality of this fit is poor and the cross section is
significantly larger than the present measurement in the mass region of the
$f'_2(1525)$.

\subsection{The structure around 2.0~GeV}

We see a broad resonant structure around $W=2.0$~GeV in the cross section. However,
there is no remarkable structure in the individual $\sigma_a$, 
$\sigma_b$ and $\sigma_c$ distributions near 2.0~GeV, nor in the
total cross section $\sigma_{\rm tot} = \sigma_a +  \sigma_b +  \sigma_c$
(wherein the interference term $s_I$ drops out).  An apparent lack of
consistency between this situation and the $W$-dependence of 
the measured cross section arises mainly from the fact that 
the variation of the parameter $\sigma_b$ dominates the feature of $\sigma_{\rm tot}$  
with its large contribution at small angles and
conceals effects of the $\sigma_c$ contribution, which is enhanced only at large angles.

  Our fit gives negative values of $\sigma_c$ in the wide $W$ region
above 1.6~GeV.
This feature might imply a non-negligible (negative)
interference contribution ($s_I$) and, therefore,
sizable values for both $\sigma_{00}$ and $\sigma_{20}$. Given
the expected dominance of only
the ${\cal H}(2,2)$ component in tensor-meson production, we still cannot
explain the relatively narrow enhancement around $|\cos \theta^*|=0$
in the angular dependence for $W$ between 1.8 and 2.1~GeV.

Thus, no conclusive result is obtained for the spin of the
structure near 2.0~GeV from our angular analysis.
However, the large $R$ value near 2.0~GeV shown in Fig.~7 indicates
that an assignment of $J=2$ is qualitatively favored for this resonance, since
both $|{\cal H}(2,0)|^2$ and $|{\cal H}(2,2)|^2$ have a peak 
at $\cos \theta^*$ = 0.

Our best fit mass and total width of the resonant structure near 2.0~GeV
match those tabulated for the $f_2(2010)$ meson~\cite{pdg}.
We obtain $\Gamma_{\gamma\gamma}{\cal B}= 
61 \pm 2 \pm 13$~eV for the assumption of $(J, \lambda)=(2,2)$.  This
result is similar in scale to that of $f_2'(1525)$ and supports the
hypothesis that this resonance is a $q\bar{q}$ meson with
a sizable $s\bar{s}$ component~\cite{muenz,bolonkin}.

\begin{table*}
\caption{Resonance parameters from the results of this study compared to 
relevant previous measurements.
$M$ and $\Gamma$ are the mass and the total width of the resonance in units of MeV/$c^2$
and MeV, respectively,
and $\Gamma_{\gamma\gamma}{\cal B}$ is the product of the two-photon decay width
and the branching fraction to the $K^+K^-$ final state for the resonance
($\Gamma_{\gamma\gamma}(R){\cal B}(R \to K^+K^-)$) in units of eV.
For the present measurement, $\Gamma_{\gamma\gamma}{\cal B}$
is derived from the $S$ values for the stated $(J,\lambda)$ hypothesis. 
For prior observations (cited in the rightmost column), 
isospin invariance is assumed for comparison with our results, and any additional notes 
are given at the bottom of the table.}

\begin{center}
\begin{tabular}{crl|crlc} 
\hline
Reson- &\multicolumn{2}{c|}{The present } &\multicolumn{4}{c}{Other possibly }\\
ance&\multicolumn{2}{c|} {measurement } &\multicolumn{4}{c}{relevant observations}\\
\hline
$f_2'(1525)$ & $M$: &$1518 \pm 1 \pm 3$ & $f_2'(1525)$ & $M$:& $1525 \pm 5$ & \cite{pdg}\\
             & $\Gamma$: & $82 \pm 2 \pm 3$ &  & $\Gamma$:& $76\pm 10$ & \\
             & $\Gamma_{\gamma\gamma}{\cal B}$: & $28.2 \pm 2.4 \pm 5.8$ & & $\Gamma_{\gamma\gamma}{\cal B}$:& $40 \pm 4.5$ & \\
             &           & \ ((2,2) assumed)     & & $\Gamma_{\gamma\gamma}{\cal B}$:& $15.7 \pm 2.5 \pm 3.9^{1)}$ & \cite{cKK}\\
\hline
1.75~GeV     & $M$:& $1737 \pm 5 \pm 7$ & $f_J(1710)^{2)}$ &  $M$:& $1767 \pm 14$ & \cite{L3}\\
             & $\Gamma$:& $ 151 \pm 22 \pm 24 $ &  & $\Gamma$:& $187 \pm 60$ & \\
             & $\Gamma_{\gamma\gamma}{\cal B}$:& $10.3 \pm 2.1 \pm 2.3$ & & $\Gamma_{\gamma\gamma}{\cal B}$:& $24.5 \pm 5.5 \pm 6.5$ & \\
             &                              &  \ ((2,2) assumed)  & $f_0(1710)$ & $M$:& $1713 \pm 6$ & \cite{pdg} \\
             &                & $76 \pm 15 \pm 17$ &                &     $\Gamma$:& $125 \pm 10$ & \\
             &                & \ ((0,0) assumed)  &    $a_2(1750)$ & $M$:& $1752 \pm 21 \pm 4$ & \cite{L33pi} \\
             &                & &                & $\Gamma$:& $ 150 \pm 110 \pm 34$ & \\
             &                & & $a_2(1700)$ & $M$:& $1726 \pm 26$ & \cite{pdg} \\
             &                & &             & $\Gamma$:& $ 256 \pm 40$ & \\
\hline
2.0~GeV      & $M$:& $1980 \pm 2 \pm 14$    & $f_2(2010)$ & $M$:& $2011^{+62}_{-76}$ & \cite{pdg} \\
             & $\Gamma$:& $297 \pm 12 \pm 6$ &               &$\Gamma$:& $202^{+67}_{-62}$ & \\
             & $\Gamma_{\gamma\gamma}{\cal B}$:& $61 \pm 2 \pm 13$ & & & & \\
             &           & \ ((2,2) assumed)     & &  & & \\
\hline
2.3~GeV     & $M$: & $2327 \pm 9 \pm 6$ & $f_J(2220)$ &  $M$: & $2231.1 \pm 3.5$ & \cite{pdg}\\
             & $\Gamma$: & $ 275 \pm 36 \pm 20 $ &  & $\Gamma$:& $23^{+8}_{-7}$ & \\
             & $\Gamma_{\gamma\gamma}{\cal B}$:& $22  \pm 3 \pm 6$ & $f_2(2300)$ & $M$:& $2297 \pm 28$ & \cite{pdg} \\
             &     &  \ ((2,2) assumed)  &  & $\Gamma$: & $149 \pm 41$ &  \\
             &                & $161 \pm 22 \pm 48$ &  & & & \\
             &                & \ ((0,0) assumed)  & & & & \\  
\hline  
\end{tabular}
\ \\
\vspace*{3mm}
The units of $M$, $\Gamma$, and $\Gamma_{\gamma\gamma}{\cal B}$ are  
MeV/$c^2$, MeV, and eV, respectively.\\
\end{center}
\hspace*{1.5cm} $^{1)}$ Using a coherent background.\\
\hspace*{1.5cm} $^{2)}$ Spin 2 is reported to be dominant in the measurement~\cite{L3}.\\

\end{table*}

\subsection{The structures around 1.75~GeV and 2.3~GeV}

Our best fit mass and width of the resonance structure at 1.75~GeV 
in our data are compatible with the corresponding values of 
a resonance in ref.~\cite{L3} that the
authors associate with $f_J(1710)$ in their measurement of 
$\gamma \gamma \to K_S^0 K_S^0$.
These numbers are excluded from the latest world-average values~\cite{pdg} of
the $f_0(1710)$ since ref.~\cite{L3}
favors a spin assignment of $J=2$ while ref.~\cite{pdg} has assigned it
$J=0$.
The measurement of the $\pi^+\pi^-\pi^0$
final-state production from two-photon collisions~\cite{L33pi} 
also finds a resonance near 1.75~GeV with a mass and width 
close to our values (they are shown as $a_2(1750)$ in Table~4).
Ref.~\cite{pdg} cites this resonance as $a_2(1700)$, with a 
world-average value that includes ref.~\cite{L33pi} 
as well as results from hadron-beam experiments.  

  We cannot determine the spin-helicity of this structure
from the present analysis alone, 
since the fraction of its contribution to the total
cross section is small.  The isospin of the resonant 
structure also can not be determined from the present experiment only.  
However, the difference in resonant structures between the $K^+K^-$  
and $K^0_SK^0_S$ channels, where in the latter process
no enhancement 
near 2.0~GeV is seen~\cite{L3}, can be naively explained 
by distinct interference effects of two or more
resonances having different isospins.  Along this line,
interference between $I=0$ and $I=1$ mesons---$f$ and $a$ mesons,
respectively---is a plausible 
explanation for the contrasting behavior between 
the $K^+K^-$  and  $K^0_SK^0_S$
channels between 1.6 and 2.1~GeV.

As noted earlier, the angular distribution
is rather flat near $W =2.3$~GeV.  However, it
is not clear that this is due to a spin property 
of the resonance  we find in this mass region, since the 
observed evolution of the angular dependence with energy may come from
the effect of hard parton scatterings,
which have a tendency of giving a
forward enhancement at high energies.

In fact, the feature of the negative $\sigma_c$ in $W > 1.6$~GeV, mentioned in the
previous subsection, may be simply an artifact of our having neglected
the partial waves of $J \geq 4$, for example, due to hard scatterings with partons,
which cause an enhancement at forward angles (corresponding to high-$J$
waves) at these higher energies.  Our fit compensates for
the omission of any $J \geq 4$ partial waves with a negative (positive)
offset in $\sigma_c$ ($\sigma_b$).
We get reasonable fits of the angular distributions for $W>2.2$~GeV with a 
sum of incoherent amplitudes, $|{\cal H}({\rm QCD})|^2 + |{\cal H}(2,2)|^2$,
where we assume $|{\cal H}({\rm QCD})|^2$ has an angular dependence proportional
to $\sin^{-4} \theta^*$~\cite{handbag}.  Then, the drastic change of the angular distribution 
observed at 2.1 - 2.3 GeV provides very important information about
the perturbative-QCD nature of this process.

 We cannot find any evidence of a narrow-width structure 
corresponding to the glueball candidate $f_J(2220)$ that is reported
in the radiative decays of 
$J/\psi$ near 2.23~GeV~\cite{Jpsi} with $\Gamma < 30$~MeV.
However, this narrow width is not well established, so it is possible
that a part of the wide structure that we see in this mass region comes from 
the $f_J(2220)$. We note that the identity of  $f_J(2220)$ 
in relation with other resonances in the same
mass region---$f_2(2300)$ and $f_4(2300)$, for example---is not yet clarified.

Our size parameter $S$ for the resonant structure gives  $\Gamma_{\gamma\gamma}{\cal B} 
= 22 \pm 3 \pm 6$~eV for the spin-helicity hypothesis of
$(J,\lambda)=(2,2)$. This does not contradict 
the upper limit value for $f_J(2220)$ obtained from previous 
two-photon
measurements --- for example, the 95\% C.L. upper limit for the
$K^0_SK^0_S$ channel~\cite{cleonew} is 
$\Gamma_{\gamma\gamma}(f_J(2220)){\cal B}(f_J(2220) \to K^0_SK^0_S) <
1.0$~eV for a pure $(2,2)$ assumption---since those analyses assume a narrow
width of $\Gamma(f_J(2220))= 20$ - 30~MeV.  Our analysis prefers a width
about ten times larger.

Working with the hypothesis of a narrow-width $f_J(2220)$, we extract an upper
limit for its $\Gamma_{\gamma\gamma}{\cal B}$  by fixing its
mass and width at 2231~MeV/$c^2$ and 23~MeV, respectively~\cite{pdg},
and assuming pure $(J,\lambda)=(2,2)$ production.
We fit the invariant-mass distribution of the signal events in
$|\cos \theta^*|<0.6$ with a sum of a second order polynomial
plus a Breit-Wigner function in the mass range between 2.13 and 2.33
GeV/$c^2$.
We see no significant excess from the smooth polynomial level.
Our 95\% C.L. upper limit is
$\Gamma_{\gamma\gamma}(f_J(2220)){\cal B}(f_J(2220) \to K^+K^-) <
0.60$~eV,  corresponding to a fit including this resonance whose $\chi^2$
exceeds that of the best fit by $(1.64)^2$. 
We account for the systematic error in the measurement by
inflating the upper limit by $1\sigma$ of the total systematic error.
This limit is an improvement on the value from the $K^0_SK^0_S$ measurement
which is cited from ref.~\cite{cleonew} in the previous paragraph, assuming we can
compare the two by isospin invariance.

\section{Conclusion}
  The production of $K^+K^-$ in two-photon collisions has been studied using
a 67~fb$^{-1}$ data sample.
The $\gamma\gamma$ center of mass energy dependence of the cross section for 
the process $\gamma\gamma \to K^+K^-$ is measured in the range
1.4 - 2.4~GeV with much higher statistical precision than was achieved
previously.

A clear peak for $f'_2(1525)$ is observed, as reported by prior
experiments.  We find three new resonant structures in the vicinities
of 1.75~GeV, 2.0~GeV and 2.3~GeV, respectively. The mass, width and size
parameters for these resonant structures have been obtained from the fit
of the energy dependence of the cross section.

The structure around 2.0~GeV has sizeable contribution to the cross section,
and the angular distribution has a large-angle enhancement in this mass region.

The angular dependence of the differential cross section has a drastic change
at 2.1 - 2.3~GeV; below this energy, the differential cross section is more
enhanced at large angles, and it is more enhanced at small angles above this
energy. 

We do not find any signature of a narrow ($\Gamma < 30$~MeV) structure
in the vicinity of 2.23~GeV.

 We hope some questions raised by  this paper will be clarified by 
further data processing as well as combined analysis of 
different final states.
On the other hand, better phenomenological models taking into
account the interfering resonances together with QCD 
nonresonant contributions are certainly needed.\\
\ \\
\ \\

\noindent
{\bf Acknowledgements}\\
We wish to thank the KEKB accelerator group for the excellent
operation of the KEKB accelerator.
We acknowledge support from the Ministry of Education,
Culture, Sports, Science, and Technology of Japan
and the Japan Society for the Promotion of Science;
the Australian Research Council
and the Australian Department of Education, Science and Training;
the National Science Foundation of China under contract No.~10175071;
the Department of Science and Technology of India;
the BK21 program of the Ministry of Education of Korea
and the CHEP SRC program of the Korea Science and Engineering Foundation;
the Polish State Committee for Scientific Research
under contract No.~2P03B 01324;
the Ministry of Science and Technology of the Russian Federation;
the Ministry of Education, Science and Sport of the Republic of Slovenia;
the National Science Council and the Ministry of Education of Taiwan;
and the U.S.\ Department of Energy.

\end{document}